\documentclass[sigconf]{acmart}
\AtBeginDocument{%
  \providecommand\BibTeX{{%
    \normalfont B\kern-0.5em{\scshape i\kern-0.25em b}\kern-0.8em\TeX}}}

\copyrightyear{2021}
\acmYear{2021}
\setcopyright{acmcopyright}
\acmConference[ICMR '21]{Proceedings of the 2021
International Conference on Multimedia Retrieval}{August 21--24, 2021}{Taipei,
Taiwan}
\acmBooktitle{Proceedings of the 2021 International Conference on Multimedia
Retrieval (ICMR '21), August 21--24, 2021, Taipei, Taiwan}
\acmPrice{15.00}
\acmDOI{10.1145/3460426.3463619}
\acmISBN{978-1-4503-8463-6/21/08}

\usepackage{comment}
\usepackage{bm}
\usepackage{multirow}
\newcommand{\tabincell}[2]{\begin{tabular}{@{}#1@{}}#2\end{tabular}}

\acmSubmissionID{icmr070}

\begin{document}

\title{MS-SincResNet: Joint learning of 1D and 2D kernels using multi-scale SincNet and ResNet for music genre classification}

\author{Pei-Chun Chang}
\affiliation{%
  \institution{Department of Computer Science, National Yang Ming Chiao Tung University}
  \city{Hsinchu}
  \country{Taiwan}}
\email{pcchang.cs05@nycu.edu.tw}

\author{Yong-Sheng Chen}
\affiliation{%
  \institution{Department of Computer Science, National Yang Ming Chiao Tung University}
  \city{Hsinchu}
  \country{Taiwan}}
\email{yschen@nycu.edu.tw}

\author{Chang-Hsing Lee}
\affiliation{%
  \institution{Department of Computer Science Information Engineering, Chung Hua University}
  \city{Hsinchu}
  \country{Taiwan}}
\email{chlee@chu.edu.tw}


\begin{abstract}
In this study, we proposed a new end-to-end convolutional neural network, called MS-SincResNet, for music genre classification.
MS-SincResNet appends 1D multi-scale SincNet (MS-SincNet) to 2D ResNet as the first convolutional layer in an attempt to jointly learn 1D kernels and 2D kernels during the training stage.
First, an input music signal is divided into a number of fixed-duration (3 seconds in this study) music clips, and the raw waveform of each music clip is fed into 1D MS-SincNet filter learning module to obtain three-channel 2D representations.
The learned representations carry rich timbral, harmonic, and percussive characteristics comparing with spectrograms, harmonic spectrograms, percussive spectrograms and Mel-spectrograms.
ResNet is then used to extract discriminative embeddings from these 2D representations.
The spatial pyramid pooling (SPP) module is further used to enhance the feature discriminability, in terms of both time and frequency aspects, to obtain the classification label of each music clip.
Finally, the voting strategy is applied to summarize the classification results from all 3-second music clips.
In our experimental results, we demonstrate that the proposed MS-SincResNet outperforms the baseline SincNet and many well-known hand-crafted features.
Considering individual 2D representation, MS-SincResNet also yields competitive results with the state-of-the-art methods on the GTZAN dataset and the ISMIR2004 dataset.
The code is available at \url{https://github.com/PeiChunChang/MS-SincResNet}.
\end{abstract}

\ccsdesc[300]{Computing methodologies~Artificial intelligence}
\ccsdesc[500]{Computing methodologies~Neural networks}
\ccsdesc[500]{Computing methodologies~Learning latent representations}

\keywords{music genre classification, convolutional neural networks, SincNet, ResNet}


\maketitle

\section{Introduction}
\label{sec:intro}
Automatic music genre classification (MGC) is an important task for multimedia retrieval systems. 
The task of MGC is to assign a proper music genre/type to a music signal. 
Traditional MGC systems typically consist of two main stages: feature extraction and classification. 
First, some discriminative features are extracted from the input music signal, and then a classifier is used to get the music genre label according to the extracted features.
In general, short-term representation describing the timbral characteristics of music signals is first extracted from every short time window (or frame).
The most well-known timbral features include zero crossing rate (ZCR) \cite{gouyon2000use}, short time energy \cite{lu2003content}, spectral centroid/rolloff/flux \cite{rabiner1993fundamentals}, Mel-frequency cepstral coefficients (MFCC) \cite{tzanetakis2002musical}, linear prediction coefficients (LPC) \cite{tindale2004retrieval}, discrete wavelet transform (DWT) coefficients \cite{lin2005audio}, octave-based spectral contrast (OSC) coefficients \cite{jiang2002music}, MPEG-7 normalized audio spectrum envelope (NASE) \cite{kim2004audio}, etc.
Then, several short-term features extracted from consecutive frames are aggregated to form the long-term features representing the whole music signal. 
The most widely used approaches to aggregating short-term features include statistical moments \cite{tzanetakis2002musical, lidy2005evaluation, morchen2005modeling}, entropy or correlation \cite{morchen2005modeling}, nonlinear time series analysis \cite{morchen2005modeling}, autoregressive (AR) models or multivariate autoregressive (MAR) models \cite{meng2007temporal}, modulation spectral analysis \cite{lidy2005evaluation, morchen2005modeling, lee2007automatic, lee2009automatic}, bag of words (BoW) model \cite{vaizman2014codebook, su2014systematic}, vector of locally aggregated descriptors (VLAD) \cite{mironicua2016modified, liu2015making}, etc.
Given the extracted features representing the music signal, a number of supervised or unsupervised classification approaches have been developed for music genre classification \cite{aucouturier2003representing, gouyon2000use, rabiner1993fundamentals, tzanetakis2002musical}, including support vector machines (SVM) \cite{kour2015music, huang2012movie}, Gaussian mixture model (GMM) \cite{kaur2017study}, principal component analysis (PCA) \cite{jin2006random}, linear discriminant analysis (LDA) \cite{li2003comparative}, etc.

\begin{figure*}[t]
    \centering
    \includegraphics[clip,trim= 0cm 13cm 1cm 3cm, width=\textwidth]{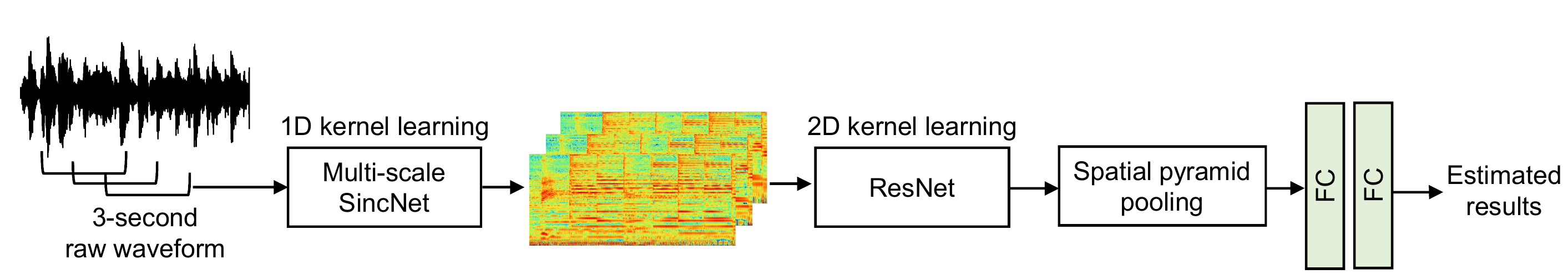}
    \caption{The proposed MS-SincResNet architecture for music genre classification. The audio waveform is first resampled to 16 kHz and divided into several overlapped 3-second music clips with each clip being fed into the proposed network to get the classification label. In the 1D kernel learning stage, the multi-scale SincNet is designed to learn the 2D representations. Next, the learned 2D representations are fed into 2D kernel learning module (the ResNet-18 is used in this study) for extracting embeddings across time and frequency aspect, simultaneously. Then, spatial pyramid pooling module is used to get the compact features from the output of the last convolutional layer (i.e., conv5\_2) of the ResNet. Finally, two fully-connected layers are applied to obtain the music genre label. In this study, the classification results of all 3-second music clips segmented from the same song will be summarized by the voting strategy to get the final classification label.}
    \label{fig:arch}
\end{figure*}

In recent years, with the remarkable success of deep learning techniques in computer vision applications, deep neural networks (DNNs) have also shown great success in speech/music classification or recognition tasks, such as speaker recognition \cite{muckenhirn2018towards, ravanelli2018speaker}, music genre classification \cite{bian2019audio, ng2020multi}, speech emotion recognition \cite{trigeorgis2016adieu}, etc. 
In these tasks, deep learning provides a new way to extract discriminative embeddings from those famous hand-crafted acoustic features, called i-vector content, for classification/recognition purposes.
To this end, deep learning methods based on convolutional neural networks (CNNs) are the most widely used approach to obtain embeddings from those i-vector content, such as MFCC \cite{tang2018music, thiruvengatanadhan2018music, vishnupriya2018automatic}, OSC coefficients \cite{vogler2016music}, 2D representations like audio spectrogram or chromagram \cite{bian2019audio, ng2020multi}, etc.

Bisharad et al. proposed a music genre classification system using residual neural network (ResNet) based model \cite{bisharad2019music2}. 
Specifically, ResNet-18 is used to extract time-frequency features from the Mel-spectrogram of each 3-second music clip. 
By taking the advantage of recurrent neural network (RNN) on sequential data analysis, they also proposed a CNN with gated recurrent unit (GRU) for music genre recognition \cite{bisharad2019music}. 
First, they apply CNN on the Mel-spectrogram to get the embedding of each 3-second music clip, and then apply RNN on the successive time-aligned embeddings for music genre classification. 
Their experiments have shown that Mel-spectrograms are capable of providing consistent performance on the GTZAN and the MagnaTagATune datasets. 

Ng et al. \cite{ng2020multi} proposed the FusionNet to combine the classification results obtained from a set of hand-crafted features, including timbre, rhythm, Mel-spectrogram, constant-Q spectrogram \cite{holighaus2012framework}, harmonic spectrogram \cite{driedger2014extending}, percussive spectrogram\cite{driedger2014extending}, scatter transform spectrogram \cite{anden2014deep}, and transfer feature\cite{choi2017transfer}.
They fed each feature into the individual feature coding network with NetVLAD \cite{arandjelovic2016netvlad} and self-attention to obtain the classification results.
Finally, they try all possible combinations among these 8 features using sum rule to report the highest testing accuracy.

Rather than applying 2D CNN on variant hand-crafted spectrograms, some researchers try to directly apply 1D CNN on the waveforms of speech/music signal to learn acoustic features \cite{kim2018sample, park2020cnn}.
In these standard 1D CNN architectures, the learnable parameters are the kernel/filter coefficients.
Typically, kernels with a large number of coefficients are needed in order to effectively characterize the timbral/rhythmic properties of the music signal, which often takes a large amount of computational cost during the training process.
To reduce the number of learnable parameters, Ravanelli et al. proposed a new architecture, called SincNet, in which a set of SincNet filters is appended to the CNN structure as the first convolutional layer \cite{ravanelli2018speaker}.
In fact, the SincNet filters are the inverse Fourier transform of some rectangular (ideal) band-pass filters parameterized with the cut-off frequencies of Mel-scale band-pass filters.
That is, these SincNet filters, can be viewed as 1D kernels used for performing 1D convolutions on the raw waveform.
Their experiments have shown that the learned SincNet filters can extract features like customized  band-pass filters with faster convergence, fewer parameters, and interpretable kernels.

As stated above, the main streams of MGC approaches typically used 2D CNN with hand-crafted 2D represnetations (for example, spectrogram, Mel-spectrogram, harmonic spectrogram, percussive spectrogram, etc.), as input.
We conjecture that if the input 2D representations can be learned using the training data, a better classification accuracy can be obtained.
Based on this idea, we propose a new end-to-end CNN architecture, called MS-SincResNet, which can jointly learn both 1D kernels and 2D kernels for MGC task.
In the proposed network architecture, 1D multi-scale SincNet (MS-SincNet) filters are appended to 2D ResNet structure as the first convolutional layer.
Given 1D raw waveform as input, MS-SincNet tries to learn variant 2D representations having different frequency resolutions. 
These learned 2D representations are stacked and then fed into 2D ResNet, followed by a spatial pyramid pooling (SPP) module to extract discriminative features for music genre classification.
Our main contribution can be summarized as follows: (1) a multi-scale SincNet (MS-SincNet) is designed to learn 2D representations from 1D raw waveform signal; (2) a new network architecture, called MS-SincResNet, which can jointly learn 1D kernels and 2D kernels, is proposed for MGC purpose.

\section{The proposed method}
As shown in Fig. ~\ref{fig:arch}, an end-to-end CNN architecture is designed to jointly learn 1D kernels and 2D kernels for music genre classification.
First, each input music signal is resampled to 16 kHz and divided into fixed-duration (3 seconds) music clips with hop size 0.5 seconds. 
Each music clip will then be fed into the proposed CNN model for MGC purpose. 
The classification results of all music clips will be summarized by using voting strategy to get the classification label for the input music signal.
For each music clip, we first exploit multi-scale SincNet (MS-SincNet) with different kernel lengths to learn variant 2D representations. 
Then, 2D ResNet and SPP module are used to extract discriminative features, followed by two fully-connected (FC) layers to obtain the music genre label from the learned 2D representations.
Before describing the proposed method, we first give a short review of the original SincNet architecture proposed by Ravanelli et al. \cite{ravanelli2018speaker}.

\subsection{SincNet}
SincNet tries to discover interpretable and meaningful filters by introducing an additional 1D convolutional layer realized by sinc-functions, followed by standard CNN layers. 
In general, it is straightforward to use rectangular (ideal) band-pass filters to decompose a signal into a number of frequency bands in the frequency domain. 
In fact, the frequency response of a band-pass filter can be written as the difference of two rectangular low-pass filters:
\begin{equation}
    G_{f_1, f_2}(f) = rect \left( \frac{f}{2f_2} \right) - rect \left( \frac{f}{2f_1} \right)
\end{equation}
where $f_1$ and $f_2$ ($f_2>f_1$) are the low and high cut-off frequencies, and $rect(\cdot)$ is the frequency response of the rectangular low-pass filter defined as follows:
\begin{equation}
rect(x)=
\left\{
    \begin{array}{lr}
    0, & \textrm{if $|x|>0.5$},  \\
    0.5, & \textrm{if $|x|=0.5$},\\
    1, & \textrm{if $|x|<0.5$}, 
    \end{array}
\right.
\end{equation}
By performing inverse Fourier transform on the filter function $G$, we can get the impulse response of the filter, represented by the sinc function:
\begin{equation}
    g_{f_1, f_2}[n] = 2f_2sinc(2\pi f_2n) - 2f_1sinc(2\pi f_1n), n = 1, 2, ..., L.
\end{equation}
where $L$ is the filter length, the sinc function is defined as $sinc(x)=\sin(x)/x$. 
In general, the sinc function is multiplied by a window function to smooth out the abrupt discontinuities of the sinc function:
\begin{equation}
    g_{f_1, f_2}^w[n] = g_{f_1, f_2}[n] \cdot w[n]
\end{equation}
In their study, the Hamming window is used, defined as follows:
\begin{equation}
w[n] = 0.54 - 0.46 \cdot \cos \left( \frac{2\pi n}{L} \right)
\end{equation}

In the original SincNet architecture, the first convolution layer consists of 80 SincNet filters of lenght $L$=251 (i.e., each filter consists of 251 coefficients), followed by two standard convolutional layers with 60 filters of length 5.
Layer normalization \cite{ba2016layer} was applied to the input raw waveform and the output of each convolution layer.
Finally, the classifier consists of three fully-connected layers having 2048 neurons, followed by batch normalization.
To increase non-linearity, all the hidden layers are followed by a Leaky-ReLU activation function.
The parameters of the SincNet filter were initialized with the cut-off frequencies derived from the Mel-scale decomposition.

\begin{figure}[t]
    \centering
    \includegraphics[clip,trim= 0cm 11cm 8cm 3cm, width=0.47\textwidth]{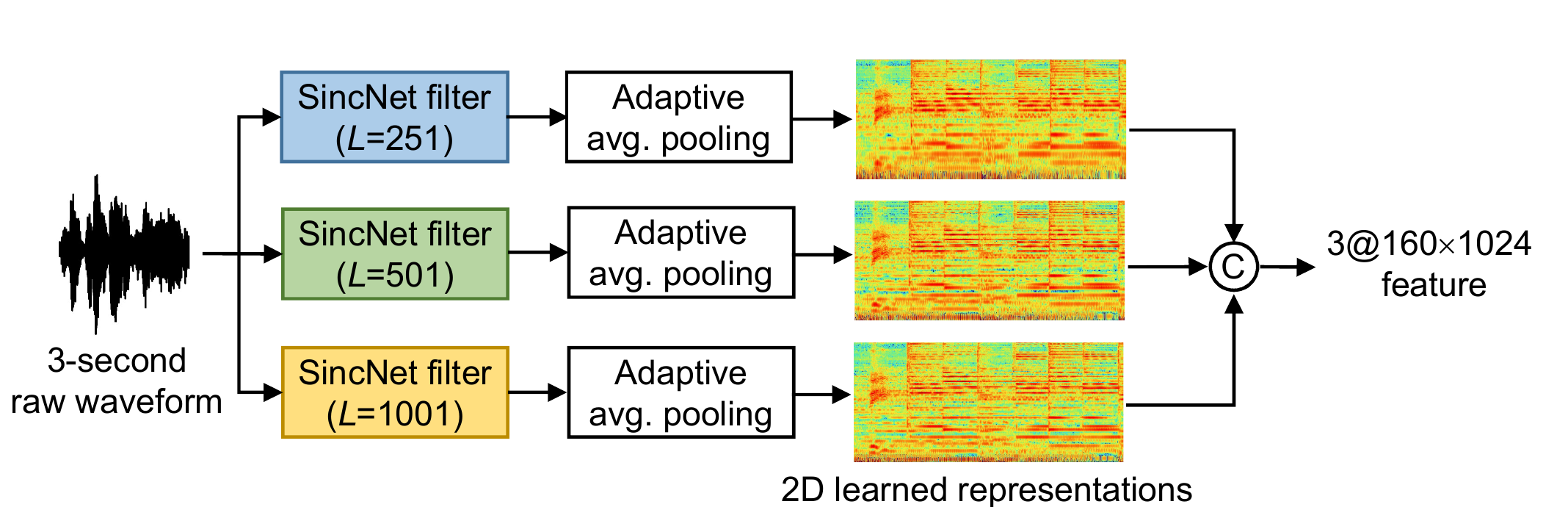}
    \caption{Illustration of 1D kernel learning using multi-scale SincNet filter. Three sets of SincNet filters with differnet filter lengths ($L$=251, 501, and 1001) are first used to learn three 2D representations having different frequency resolutions. The adaptive average pooling is then used to get compact 2D representations. Finally, we stack these compact 2D representations to obtain a three-channel image (3@160$\times$1024) which will be fed into ResNet for extracting embeddings.}
    \label{fig:SincNet_filter}
\end{figure}

\begin{figure}[t]
    \centering
    \includegraphics[clip,trim= 0.45cm 12cm 13cm 3.5cm, width=0.45\textwidth]{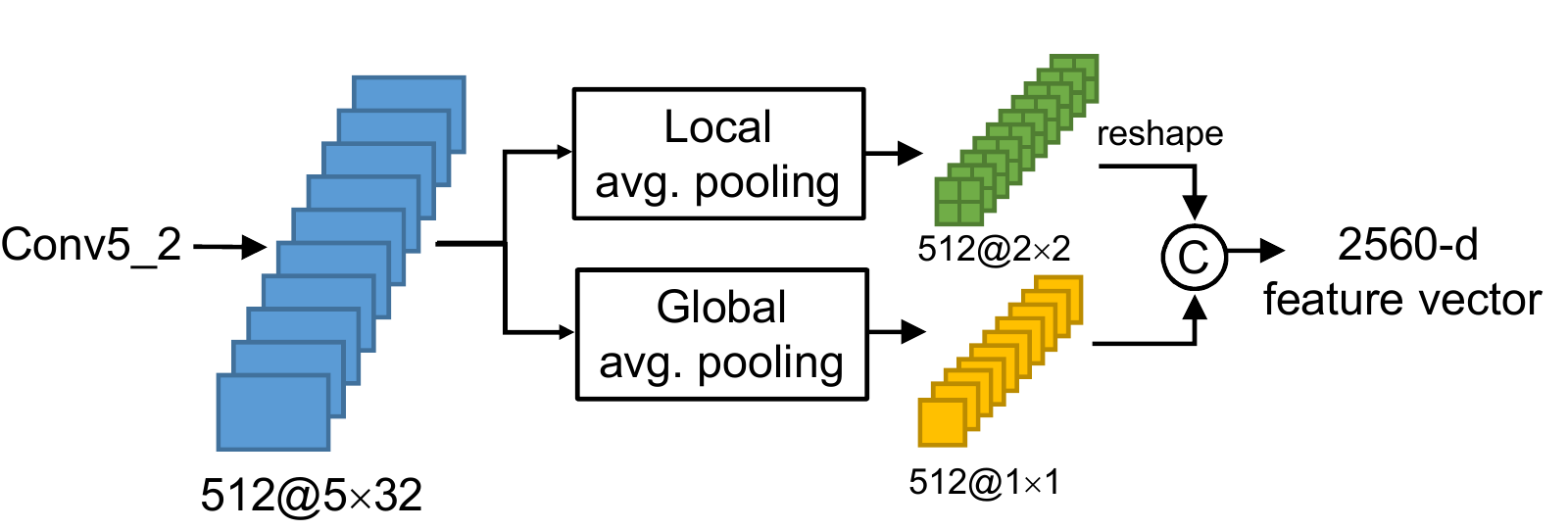}
    \caption{Illustration of spatial pyramid pooling module. The input is 512@5$\times$32 feature volume from the output of the last convolutional layer of ResNet-18. There are two branch to aggregate both local and global features using average pooling on blocks of different sizes. After that, we flattened and concatenated these pooled values as a 2560-d compact feature vector.}
    \label{fig:spp}
\end{figure}

In standard 1D CNN \cite{kim2018sample, park2020cnn}, the number of learnable parameters for each filter is $L$, the filter length.
However, for each SincNet filter, there are only two parameters ($f_1$ and $f_2$), which represent respectively the low and high cut-off frequencies of the band-pass filters, have to be learned during the training process. 
Therefore, comparing with standard 1D CNN, SincNet can obtain faster convergence using fewer parameters, and provide interpretability of the neural networks \cite{ravanelli2018speaker}.


\subsection{Proposed Multi-scale SincResNet (MS-SincResNet)}
Given the raw waveform of a music clip, multi-scale SincNet (MS-SincNet), which consists of three sets of SincNet filters of different lengths, is first designed to learn 2D representations.
The outputs of each set of SincNet filters are concatenated to form a 2D representation, and then all 2D representations are stacked and fed into the ResNet for extracting embeddings.
Finally, a spatial pyramid pooling module is applied to summarize the information across time and frequency aspects to obtain compact features for MGC task.
In this study, the end-to-end training strategy is used to fine-tune the parameters in the proposed MS-SincNet and ResNet.
The parameters of MS-SincNet are initialized by Mel-scale cut-off frequencies, and the parameters of ResNet are initialized by pretrained model on ImageNet dataset.

\subsubsection{Data preprocessing}
In this study, the input music signal is first resampled to 16 kHz and divided into 3-second music clips with hop size being 0.5 seconds.
Hence, each clip consists of 48,000 samples.
During the training process, two data augmentation methods, which will be described later, are applied to these music clips.
Before being fed into the proposed MS-SincResNet for embeddings extraction, each input audio waveform is normalized using layer normalization operation \cite{ba2016layer}.

\subsubsection{1D kernel learning stage}
As shown in Fig.\ref{fig:SincNet_filter}, the convolution operation in the 1D MS-SincNet filter learning stage can be represented as:
\begin{equation}
\label{eqn:conv}
    x^s_k[n] = x[n] * g^s_k[n], k = 1, 2, ..., K, s = 1, 2, 3
\end{equation}
where $K$ is the number of kernels, $x[n]$ is an input music clip having $N$ samples($N$=48,000), and $g^s_k[n]$ is the $k$-th convolutional kernel for scale $s$, called SincNet filter in this study.
To get a compact representation, we apply adaptive average pooling to each filter output ($x^s_k[n] \in \bm{R}^{1 \times N}$) of (\ref{eqn:conv}) to obtain 1024-d representation $m^s_k[n] \in \bm{R}^{1 \times 1024}$:
\begin{equation}
    m^s_k[n] = adaptiveAvgPool(x^s_k[n])
\end{equation}

Then, for each scale, we concatenate the compact outputs of all kernels to get the corresponding 2D representation of the music clip, $M^s \in \bm{R}^{K \times 1024}$:
\begin{equation}
M^s = 
    \begin{bmatrix}
    m^s_1[n]]\\
    m^s_2[n]\\
    \vdots \\
    m^s_K[n]\\
    \end{bmatrix}
    =
    \begin{bmatrix}
    m^s_1[1], m^s_1[2], \hdots, m^s_1[1024]\\
    m^s_2[1], m^s_2[2], \hdots, m^s_2[1024]\\
    \vdots \\
    m^s_K[1], m^s_K[2], \hdots, m^s_K[1024]\\
    \end{bmatrix}
\end{equation}
In this study, the set of parameters ($f_1$ and $f_2$) of the SincNet filters were initialized using Mel-scale cut-off frequencies between [30, $f_s / 2$] Hz, where $f_s$ is the sampling frequency.
Specifically, the values of the lower cut-off frequency $f_1$ and higher cut-off frequency $f_2$ were initialized according to the Mel-scale cut-off frequencies.
As a result, all the derived 2D representations $M^s \in \bm{R}^{K \times 1024}$ ($s$=1, 2, 3) can be considered as the learned multi-scale Mel-spectrograms of the input music clip. 

For each set of SincNet filters, 160 convolutional kernels (i.e., $K$=160) followed by a 1D batch normalization and a ReLU non-linear activation function is applied to get the filter output.
For MS-SincNet, three different kernel lengths ($L$ = 251, 501, and 1001) corresponding to three scales ($s$=1, 2, 3) are applied to the input waveform to get three 2D representations.
By stacking these 2D representations, we can obtain a $3\times160\times1024$ 2D representation for each music clip.

\subsubsection{2D kernel learning stage}
ResNet is one of the most well-known backbone network in deep neural networks \cite{he2016deep}.
Comparing to prior network architectures, ResNet introduces a shortcut connection to address the problem of vanishing gradient, and further extracts abundant sementics from the input data to build a robust classifier.
In this paper, ResNet-18 pretrained using the ImageNet dataset is selected as our 2D kernel learning backbone network.
We performed transfer learning on ResNet-18 to fine-tune the kernel parameters using music clips derived from the training set.
The input to ResNet-18 is the three-channel 2D representations $M^s$ ($s$= 1, 2, 3) obtained from the 1D MS-SincNet learning stage, and the output of the last convolution layer (i.e., conv5\_2) is a 512-channel $5\times32$ feature volume from which discriminative features will be extracted.

\subsubsection{Spatial pyramid pooling}
The spatial pyramid pooling (SPP) module is used to enhance the feature discriminability in terms of both time and frequency aspects \cite{he2015spatial}, as shown in Fig. \ref{fig:spp}.
By performing global average pooling on each channel or each block obtained by dividing the channel into 2$\times$2 blocks, we can get the SPP features consisting of 512@1$\times$1 (global feature) and 512@2$\times$2 (local feature).
Then, we flattened and concatenated all the feature values, and fed them into two fully-connected layers to obtain the classification genre label.

\subsection{Training strategy and data augmentation}
In this study, the proposed MS-SincResNet architecture, including 1D kernel and 2D kernel learning, is implemented on the Pytorch framework.
The classification result is obtained for each music clip in the training stage, whereas in the testing stage the voting strategy is used to get the final classification label of the input music signal consisting of several music clips.

The SGD optimizer is used to tune both 1D MS-SincNet parameters and 2D ResNet parameters in the whole network model.
In the training stage, the warm-up strategy with learning rate $lr=10^{-5}$ is used for the first five epochs.
After that, we set $lr$=0.005 from the 6-th epoch, and decay half for every 30 epochs.

To avoid overfitting, for each epoch we randomly select 4 music clips from each input music data to train the network.
In addition, two data augmentation methods are used to enhance the variation of the training data.
First, we multiply the amplitude of the music signal by a ratio randomly chosen within the interval [0.9, 1.1].
Second, we add zero-mean Gaussian noise ($\sigma$=0.02) to the signal.

\begin{figure*}[t]
    \centering
    \includegraphics[clip,trim= 0cm 7cm 10.5cm 3cm, width=1\textwidth]{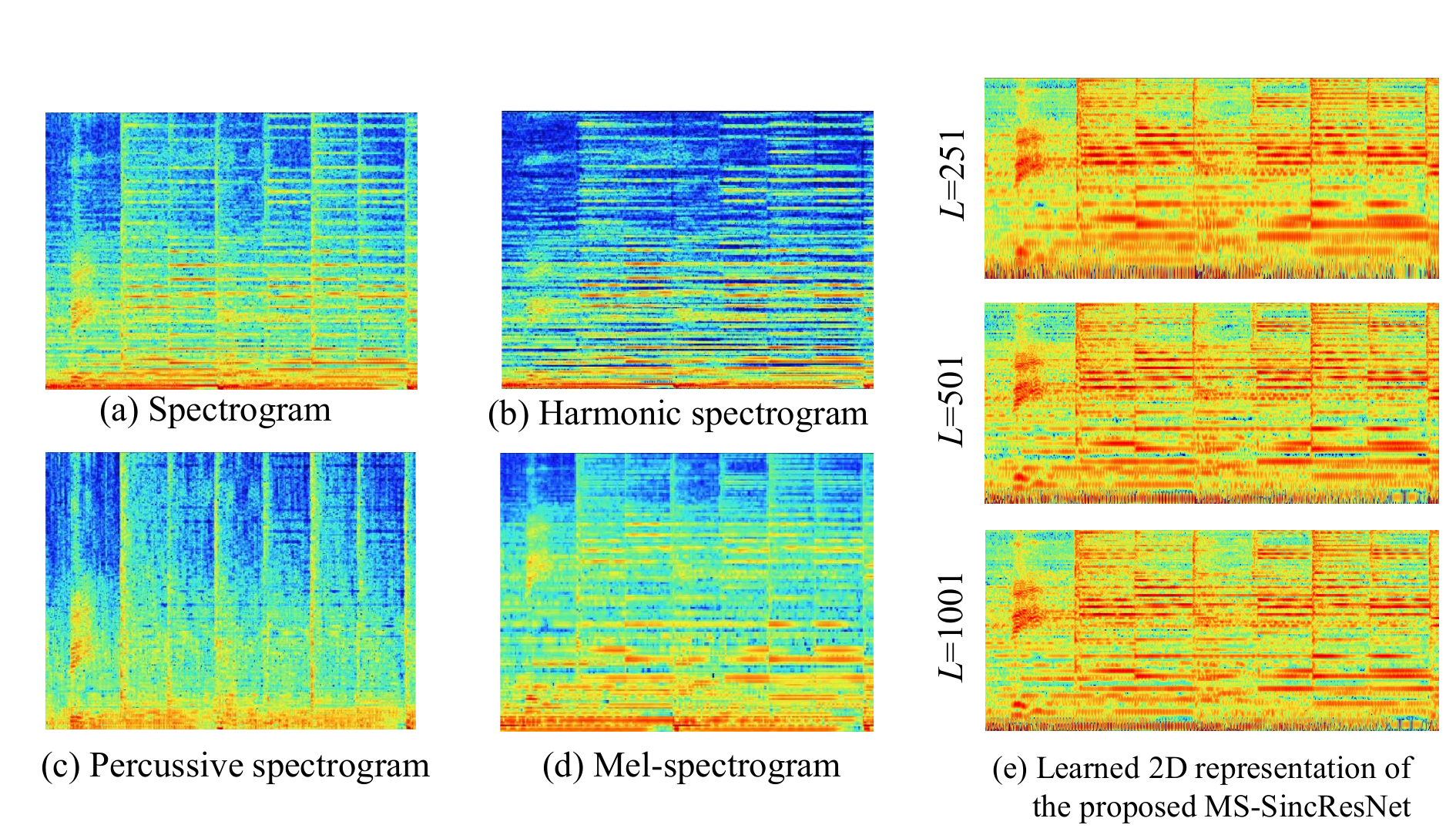}
    \caption{Visualization of different 2D representations (a) spectrogram, (b) harmonic spectrogram (c) percussive spectrogram, (d) Mel-spectrogram, and (e) 2D representations learned by the proposed MS-SincResNet.}
    \label{fig:twodrepres}
\end{figure*}

\begin{table*}[t]
    \caption{The ablation study of proposed method for SincNet filter initialization, filter number, filter length, spatial pyramid pooling (SPP) on the GTZAN and the ISMIR2004 datasets. The best classification accuracy on GTZAN and ISMIR2004 datasets are respectively 91.49\% and 91.91\% by using MS-SincResNet with 160 learnable filters initialized with Mel-scale decomposition and SPP module.}
    \centering
    \begin{tabular}{|l|l|c|c|c|c|c|c|}
    \hline
     & \tabincell{c}{SincNet filter \\ initialization} & \tabincell{c}{SincNet filter \\ number $(K)$} & \tabincell{c}{Multi-scale \\ SincNet filters} & \tabincell{c}{SincNet filter \\ length $(L)$ } & SPP & \textbf{GTZAN(\%)} & \textbf{ISMIR2004(\%)} \\
    \hline
    \multirow{6}{*}{SincNet} 
    & Mel-scale & 80 & - & 251 & - & 73.98 & \textbf{81.34} \\\cline{2-8}
    & Mel-scale & 80 & - & 501 & - & \textbf{76.37} & 79.70 \\\cline{2-8}
    & Mel-scale & 80 & - & 1001 & - & 74.97 & 79.01 \\\cline{2-8}
    & Mel-scale & 160 & - & 251 & - & 75.78 & 80.12 \\\cline{2-8}
    & Mel-scale & 160 & - & 501 & - & 76.08 & 71.83 \\\cline{2-8}
    & Mel-scale & 160 & - & 1001 & - & \textbf{76.38} & 78.33 \\\cline{2-8}
    \hline
    ResNet (Mel-spetrogram) & - & - & - & - & - & 85.49 & 88.34 \\
    \hline
    \multirow{6}{*}{SincResNet} 
    & Mel-scale & 80 & - & (251, 251, 251) & - & 89.79 & 88.75 \\\cline{2-8}
    & Mel-scale & 80 & - & (501, 501, 501) & - & 89.78 & 87.93 \\ \cline{2-8}
    & Mel-scale & 80 & - & (1001, 1001, 1001) & - & 89.69 & 86.56 \\ \cline{2-8}
    & Mel-scale & 160 & - & (251, 251, 251) & - & 90.89 & 89.99 \\ \cline{2-8}
    & Mel-scale & 160 & - & (501, 501, 501) & - & 90.79 & 90.12 \\ \cline{2-8}
    & Mel-scale & 160 & - & (1001, 1001, 1001) & - & \textbf{91.08} & \textbf{90.26} \\ 
    \hline
    \multirow{4}{*}{MS-SincResNet} 
    & Mel-scale & 80 & $\surd$ & (251, 501, 1001) & - & 90.18 & 87.24 \\ \cline{2-8}
    & Mel-scale & 80 & $\surd$ & (251, 501, 1001) & $\surd$ & 90.38 & 87.52 \\ \cline{2-8}
    & Mel-scale & 160 & $\surd$ & (251, 501, 1001) & - & 91.29 & 89.71 \\ \cline{2-8}
    & Mel-scale & 160 & $\surd$ & (251, 501, 1001) & $\surd$ & \textbf{91.49} & \textbf{91.91} \\ 
    \hline
    \end{tabular}
    \label{tab:ablation}
\end{table*}

\section{Experiments}
In this study, the GTZAN dataset \cite{GTZAN} and the ISMIR2004 Audio Description Contest dataset \cite{ISMIR} were used for performance evaluation.
In this section, we will give brief descriptions for these two datasets, and then give the experimental results.

\subsection{The GTZAN dataset}
The GTZAN dataset consists of 1000 audio tracks involved ten music categories: Blues, Classical, Country, Disco, Hip Hop, Jazz, Metal, Popular, Reggae, and Rock.
For each music genre, there are exactly 100 tracks, and all tracks were recorded in 22,050 Hz with mono 16-bit wav format.
Similar to prior works, we used 10-fold cross-validation on the GTZAN dataset to evaluate the classification performance.
For each fold, 900 tracks were randomly selected as training set, and the remaining 100 tracks were used for testing.
The performance will be computed by averaging the classification results of these 10 folds.

\subsection{The ISMIR2004 dataset}
The ISMIR2004 dataset consists of 1458 music tracks, in which 729 music tracks are used for training and the others for testing.
These music tracks are classified into six classes, including Classical, Electronic, Jazz/Blue, Metal/Punk, Rock/Pop, and World.
The audio file format is 44.1 kHz, 16 bits per sample, and stereo MP3 format. 
Among the 729 music tracks used for training, we randomly selected 1/10 as validation set in an attempt to choose the best parameter set for evaluating the performance of the testing data set.

\subsection{Baseline setups}
In this study, the original SincNet architecture with filter lengths ($L$=251/501/1001), initialized using Mel-scale cut-off frequencies, is used as the baseline network.
For the proposed SincResNet and MS-SincResNet, the setting of the SincNet filters also follows the original design \cite{ravanelli2018speaker}.
In addition, we compared MS-SincNet having different filter lengths (MS-SincResNet) with the original single scale SincNet filters followed by the ResNet architecture (SincResNet).
To show the learning capability of the proposed MS-SincNet, 2D representations learned using MS-SincNet and hand-crafted Mel-spectrogram are individually fed into ResNet to compare their classification accuracy.


\section{Experimental results}
First, we compare the visualization of the 2D representations learned using MS-SincResNet with other 2D representations.
Then, we show the ablation study to investigate the performance of the proposed MS-SincResNet approach.
Finally, we compare the proposed MS-SincResNet approach with other competitive approaches on the GTZAN and the ISMIR2004 datasets.

\subsection{The learned 2D representation}
In this section, we compare the 2D representations learned using MS-SincResNet with other 2D representations, such as spectrogram, harmonic spectrogram, percussive spectrogram, and Mel-spectrogram.
The spectrogram and Mel-spectrogram are obtained by short-term Fourier transform with window size 512 samples and hop size 128 samples.
The harmonic spectrogram and percussive spectrogram are obtained based on the harmonic-percussive source separation algorithm \cite{fitzgerald2010harmonic}.
Each 2D representation is obtained by feeding a music clip to MS-SincNet using variant filter lengths.
From Fig.\ref{fig:twodrepres}, we can see that 2D representations learned using the proposed MS-SincNet filters exhibit noticeable harmonic-related and percussive-related features, particularly for high frequency components.
Thus, it is expected that using the learned 2D representations as input to ResNet will yield better classification accuracy than Mel-spectrogram input.

\begin{figure}[t]
    \centering
    \includegraphics[clip,trim= 4cm 8.5cm 4cm 9cm, width=0.42\textwidth]{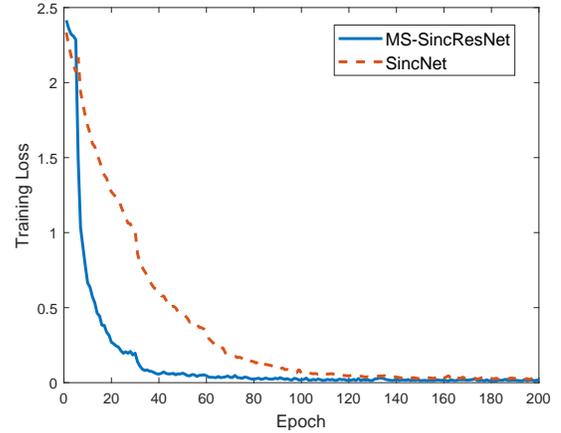}
    \caption{The training loss curve of the proposed MS-SincResNet method ($K$=160 and $L$=(251, 501, 1001)) and the original SincNet ($K$=160 and $L$=251) over various epochs.}
    \label{fig:loss}
\end{figure}

\subsection{Ablation study}
As shown in Table \ref{tab:ablation}, for baseline SincNet, the best classification accuracy on the GTZAN and the ISMIR2004 datasets are obtained by setting $K$=160, $L$=1001 (76.38\%) and $K$=80, $L$=251 (81.34\%), respectively.
That is, it is hard to select a filter length $L$ that can achieve the best classification accuracy for all datasets.
Also, we evaluated the classification results by using ResNet with Mel-Spectrogram as input.
The classification accuracy is 85.49\% and 88.34\% on the GTZAN and the ISMIR2004 datasets, respectively.
This shows that 2D ResNet with Mel-spectrogram input outperforms 1D SincNet with raw waveform input.
By replacing Mel-spectrogram with 2D representation learned using single scale SincNet (notated by SincResNet), the classification accuracy can be improved to be 91.08\% and 90.26\% when $K$=160 and $L$=1001.
This comparison shows that using SincNet to learn 2D representation can extract more discriminative features and obtain better classification accuracy than hand-crafted Mel-spectrogram features.
For MS-SincResNet, the best classification accuracy is 91.49\% and 91.91\% when SPP is incorporated in the network architecture.
Comparing with the best results obtained by the baseline SincNet, an improvement of 15.11\% and 10.57\% on the GTZAN and the ISMIR2004 dataset, respectively.

Fig.\ref{fig:loss} compares the training loss curves of the proposed MS-SincResNet with baseline SincNet on the GTZAN dataset. It demonstrates that the proposed MS-SincResNet can converge faster, and obtain better classification accuracy than baseline SincNet.

\begin{table}[t]
    \caption{Comparison the proposed method with the state-of-the-art methods on the GTZAN dataset.}
    \centering
    \begin{tabular}{|l|c|}
    \hline
    The SOTA methods & GTZAN dataset (\%)  \\ 
    \hline
    Bisharad et al. \cite{bisharad2019music} & 85.36 \\
    \hline
    Bisharad et al. \cite{bisharad2019music2} & 82.00 \\
    \hline
    Raissi et al. \cite{raissi2018extended} & 91.00 \\
    \hline
    Sugianto et al. \cite{sugianto2019voting} & 71.87 \\
    \hline
    Ashraf et al. \cite{ashraf2020globally} & 87.79 \\
    \hline
    Ng et al. \cite{ng2020multi} (FusionNet) & 96.50 \\
    \hline
    Liu et al. \cite{liu2020bottom} & 93.90 \\
    \hline
    Nanni et al. \cite{nanni2017combining} & 90.60 \\
    \hline
    Ours (MS-SincResNet) & 91.49 \\
    \hline
    \end{tabular}
    \label{tab:STOA_GTZAN}
\end{table}

\begin{table}[t]
    \caption{Comparison the proposed method with the state-of-the-art methods on the ISMIR2004 dataset.}
    \centering
    \begin{tabular}{|l|c|}
    \hline
    The SOTA methods & ISMIR2004 dataset (\%)  \\ 
    \hline
    Ng et al. \cite{ng2020multi} (FusionNet) & 92.46 \\
    \hline
    Nanni et al. \cite{nanni2017combining} & 90.90 \\
    \hline
    Nanni et al. \cite{nanni2016combining} & 90.20 \\
    \hline
    Costa et al. \cite{costa2017evaluation} & 87.10 \\
    \hline
    Ours (MS-SincResNet)  & 91.91 \\
    \hline
    \end{tabular}
    \label{tab:STOA_ISMIR}
\end{table}

\subsection{Comparison with the state-of-the-art methods}
Tables \ref{tab:STOA_GTZAN} and \ref{tab:STOA_ISMIR} compare the proposed MS-SincResNet with the state-of-the-art methods on the GTZAN and the ISMIR2004 datasets, respectively.
The classification accuracy of the proposed MS-SincResNet method is 91.49\% and 91.91\%.
From these two tables, we can see that the FusionNet proposed by Ng et al. \cite{ng2020multi} achieves the best classification accuracy.
However, as stated in Sec. \ref{sec:intro}, FusionNet tries all possible combinations among 8 different features (timbre, rhythm, Mel-spectrogram, constant-Q spectrogram \cite{holighaus2012framework}, harmonic spectrogram \cite{driedger2014extending}, percussive spectrogram\cite{driedger2014extending}, scatter transform spectrogram \cite{anden2014deep}, and transfer feature\cite{choi2017transfer}) using sum rule to get the highest testing accuracy for each dataset. 
In fact, when considering one individual network, the best performance on the GTZAN dataset is 89.10\% using Mel-spectrogram, on the ISMIR2004 dataset is 87.38\% using transfer feature.
That is, without fusion of different networks, our learned 2D representations always achieves better performance than the other hand-crafted 2D representations.

\section{Conclusions}
In this study, we proposed an end-to-end CNN architecture, called MS-SincResNet, which can jointly learn 1D kernels and 2D kernels, for music genre classification.
For 1D kernel learning, we use MS-SincNet filters to obtain variant 2D representations from raw audio waveform rather than pre-computed hand-crafted features such as Mel-spectrogram.
Then, 2D kernel learning using ResNet-18 is used to extract embeddings from these learned 2D representations.
The spatial pyramid pooling module is used to get the compact features from the output of the last convolutional layer of ResNet-18.
In the experiments, the proposed MS-SincResNet approach achieves classification accuracy of 91.49\% and 91.91\% on the GTZAN and ISMIR2004 datasets, which outperforms every hand-crafted 2D representation.

Inspired by the FusionNet \cite{ng2020multi}, we can see that the combination of the classification results of different features often yields better performance than each individual feature.
In the future, we will try to combine the classification results of variant networks in which 2D representations can be learned using SincNet with different set of cut-off frequencies as the initialization of the band-pass filters.
That is, in addition to Mel-scale decomposition, linear-scale decomposition, or other frequency decomposition approaches such as OSC and NASE can also be considered.

\begin{acks}
This work was supported in part by Ministry of Science and Technology, Taiwan (MOST-108-2221-E-216-005 and MOST-108-2221-E-009-066-MY3).

\end{acks}

\bibliographystyle{ACM-Reference-Format}
\bibliography{sample-base}

\end{document}